# Discrete wavelet structure and discrete energy of a classical plane light wave

Zhang Xingchu[1],    She Weilong[2,3*]

[1] Department of Physics and Information Engineering, Guangdong University of Education, Guangzhou 510303, China;

[2] School of Physics, Sun Yat-Sen University, Guangzhou 510275, China.

[3] Sino-French Institute of Nuclear Engineering and Technology, Sun Yat-Sen University, Zhuhai 519082, China

* Email: shewl@mail.sysu.edu.cn

In this letter, the wavelet transform is used to decompose the classical linearly polarized plane light wave into a series of discrete Morlet wavelets. It is found that the energy of the light wave can be discrete, associated with its discrete wavelet structure. It is also found that the changeable energy of a basic plane light wave packet or wave train of wave vector $\vec{k}$ and with discrete wavelet structure can be with the form of $H_{0k} = n p_{0k} \omega$ $(n=1,2,3,...)$, where $n$ is the parameter of discrete wavelet structure, $\omega$ is the idler frequency of the light wave packet or wave train, and $p_{0k}$ is a constant to be determined. This is consistent with the energy division of $P$ portions in Planck radiation theory, where $P$ is an integer. Finally, the random light wave packets with $n=1$ are used to simulate the Mach-Zehnder interference of single photons, showing the wave-particle duality of light.

One of apparitions in physics is the wave-particle duality of particles [1,2]. The property of light quantum (photon) is a typical sample [1,2,3,4]. Although there have been Dirac [5], Guputa [6] and Feimi's[7] approaches towards quantizing the radiation field and various of models presented for photon [8-13], it is very difficult to understand thoroughly the property of photon due to its multi-faceted and elusive nature [14,15]. The difficulty is mainly from the depiction of its wave feature. Unlike an electron, whose wave state can be described by a coordinates function (the probability amplitude of spatial localization), the photon with an energy of $\hbar\omega$ has no a similar probability amplitude available [16] though there exists a light wave involving the photon. Therefore, when a light wave interacts with a matter, the change of its energy is known to be discrete with basic unit of $\hbar\omega$. But it is unknown what change will happen about the wave state of light while the photons are absorbed. The following ideal experiment may be enlightened for prying into the property of photon: Let a plane light wave of frequency $\omega$ be incident onto an absolute blackbody and exam the state change of the light wave during the process of light absorption. The question then arises: what happens to the plane light wave when the photons are absorbed by absolute blackbody? Is that the amplitude of the whole incident plane wave changes constantly, or that the wave disappears segment by segment? Our experiment, using a chopped beam of He-Ne laser incident onto a photomultiplier tube in a black cavity with a small hole, shows that the absorption of photons in the front of light wave is impossible to affect the amplitude of the succeeding part of the light wave. It suggests that the plane light wave would disappear segment by segment when the photons are absorbed. Then, it also suggests that, by the time inversion, the plane light wave could be reconstructed by the translation and superposition of a segment and a segment of wave packet or wave train. Thus, another question arises, what method can do the reconstruction of light wave? The wavelet transform [17] is a candidate. In fact, the wavelet transform has been adopted in the investigation of physical problems for many years [17-19]. And Morlet wavelet function was widely used due to its advantage in simplicity and time-frequency analysis [20]. In this letter, the wavelet transform is used to decompose the classical linearly polarized plane light wave into a series of discrete Morlet wavelets (or basic wavelets). And we find that the energy of the light wave can be related to its discrete wavelet structure and therefore can be discrete as well. Finally, the random light wave packets are used to simulate the Mach-Zehnder interference, showing the "wave-particle duality" of light.

Now, we start to investigate the discrete wavelet structure and discrete energy of classical plane light waves. We first consider the radiation field in free space. As well known, the classical radiation fields in free space satisfy Maxwell's equations. And the wave equations for electric field can be derived from them as follows

$$\nabla^2 \vec{E} - \frac{1}{c^2} \frac{\partial^2 \vec{E}}{\partial t^2} = 0, \tag{1}$$

$$\nabla \cdot \vec{E} = 0, \tag{2}$$

where $\vec{E}$ is electric field vector, $c$ is the speed of light in vacuum. A plane wave solution for Eq.(1) at wavelength $\lambda$ and with wave vector $\vec{k}$ can be expressed as

$$\vec{E}_k(\vec{r},t) = q_k(t)\vec{E}_k(\vec{r}) + q_k^*(t)\vec{E}_k^*(\vec{r}), \tag{3}$$

where $q_k(t) \propto e^{-i\omega t}$, $\vec{E}_k(\vec{r}) = \vec{E}_{k0} e^{i\vec{k}\cdot\vec{r}}$, respectively, and $\vec{E}_{k0}$ is a constant vector and perpendicular to wave vector $\vec{k}$ with two independent polarized components due to Eq.(2); $q_k^*(t)$ and $\vec{E}_k^*(\vec{r})$ are conjugate complexes of $q_k(t)$ and $\vec{E}_k(\vec{r})$, respectively. The general radiation field can be expressed as the linear superposition of all possible plane waves. Now we try to decompose the electric field of plane light wave into a series of wavelets. For the sake of simplicity, we consider a linearly polarized monochromatic plane wave propagating along the z-direction, with a spatial function $E_k(\vec{r}) = E_k(z) = E_{k0}e^{i\vec{k}\cdot\vec{z}}$, where $\lambda = 2\pi/k = 500nm$.

As well known, for a given function $f(z)$, its wavelet transform and inverse wavelet transform can be expressed as [17]

$$Wf(a,b) = \frac{1}{\sqrt{a}}\int_{-\infty}^{\infty} f(z)\psi^*(\frac{z-b}{a})dz , \qquad (4)$$

$$f(z) = \frac{1}{C_\psi}\int_0^\infty da \int_{-\infty}^\infty Wf(a,b)\psi(\frac{z-b}{a})\frac{1}{a^2\sqrt{a}}db , \qquad (5)$$

where $\psi(z)$ and $\psi^*(z)$ is the basic wavelet function and its conjugate complex, respectively. $C_\psi = \int_0^\infty |\psi(\omega)|/2\pi|\omega|d\omega$, where $\psi(\omega)$ is the Fourier transform of $\psi(z)$. One sees that the original function can be reconstructed by the translation and scalation of wavelet. Here we take the simplified Morlet wavelet function for transform, which is as follows [21]:

$$\psi_K(z) = e^{iKz}e^{-z^2/(2s)} , \qquad (6)$$

where $K$ is the central wave number of the Morlet wavelet, $s$ is a parameter. For the simplified Morlet wavelet function, $C_\psi$ needs to be redefined, which will be discussed below. For the monochromatic plane wave of $K = k$ and $f(z) = E_k(z)$, its wavelet transform is

$$Wf(a,b) = \frac{E_{k0}}{\sqrt{a}}\int_{-\infty}^\infty e^{ikz}e^{-ik(\frac{z-b}{a})}e^{-(\frac{z-b}{a\sqrt{2s}})^2}dz = E_{k0}\sqrt{as}e^{ibk}e^{-s(1-a)^2k^2/2} . \qquad (7)$$

We notice that, for $E_k(z) = E_{k0}e^{i\vec{k}\cdot\vec{z}}$, the inverse wavelet transform does not need to scale and the original function can be reconstructed only by the translation. So we just take $a = 1$. Then the wavelet transform of the plane light wave can be simplified to

$$Wf(a,b) = E_{k0}\sqrt{s}e^{ibk} . \qquad (8)$$

Now we use this result to reconstruct plane light wave, according to Eq.(5),

$$E_k(z) = \frac{E_{k0}}{C_\psi}\int_{-\infty}^\infty \sqrt{s}e^{ibk}e^{ik(z-b)}e^{-(\frac{z-b}{\sqrt{2s}})^2}db = E_{k0}e^{ikz}\frac{1}{C_\psi}\int_{-\infty}^\infty \sqrt{s}e^{-\frac{(z-b)^2}{2s}}db . \qquad (9)$$

One sees that, by making $C_\psi = \int_{-\infty}^{\infty} \sqrt{s} e^{-(z-b)^2/2s} db$, the reconstruction of $E_k(z)$ is reached at once. It means that a plane light wave can be reconstructed by a series of wavelets through translation and superposition. This is the case of continuous wavelet transform. However, we are interested in that of discrete wavelet transform. The theory of discrete wavelet transform is rather complicated. Here we simplify the way for discretization of wavelet transform, according to our experience. The key is to choose a suitable parameter for $s$. We find that when $s = (\lambda/c_1)^2/2$, $c_1 = 0.886231921$, Eq. (9) can be discretized as follows

$$E_k(z) = \frac{E_{k0}}{2} e^{ikz} \lim_{n\to\infty} \sum_{r=-n}^{n+1} e^{\frac{-(z-r\lambda+\lambda/2)^2}{2s}}, \qquad (10)$$

where $n$ is the parameter of discrete wavelet structure. Equation (10) shows the discrete wavelet structure of the spatial function of plane light wave, which is the translation and superposition of infinite ones of wavelets. In order to see how to reconstruct the plane light wave, we consider the cases where n is finite. Fig.1 shows the results of wavelet reconstruction, where (a) -(d) are the cases for n = 0,1,2, and 6 with Eq. (10), respectively. The dotted line in the figure is the waveform of plane light wave. And the solid lines are those of reconstructed light wave packets or light wave trains with discrete wavelet structure. One sees that there is not a complete period for the cases of n<2. But when n=2, there are two complete periods. Every time when the value of n increases by 1, two complete periods will be added. When n=6, there are 10 complete periods. The larger n is, the more complete periods are formed by reconstruction. For n>2, the parts of incomplete period just make a wave packet of n = 1, which contains four basic wavelets. These results will be used below. It is easy to test, if n is finite in Eq.(10), making the transformation $z \to z - ct$, then we have

$$E_k(z-ct) = \frac{E_{k0}}{2} e^{ik(z-ct)} \sum_{r=-n}^{n+1} e^{\frac{-[(z-ct)-r\lambda+\lambda/2]^2}{2s}}, \qquad (11)$$

which is also the solution to Eqs.(1) and (2), representing a light wave packet or wave train, where $\lambda = 2\pi/k$ (therefore $kc = \omega$ is the idler frequency). So Fig.1 can be seen as the waveforms of light wave packets or wave trains reconstructed by wavelets at t = 0.

4 / 10

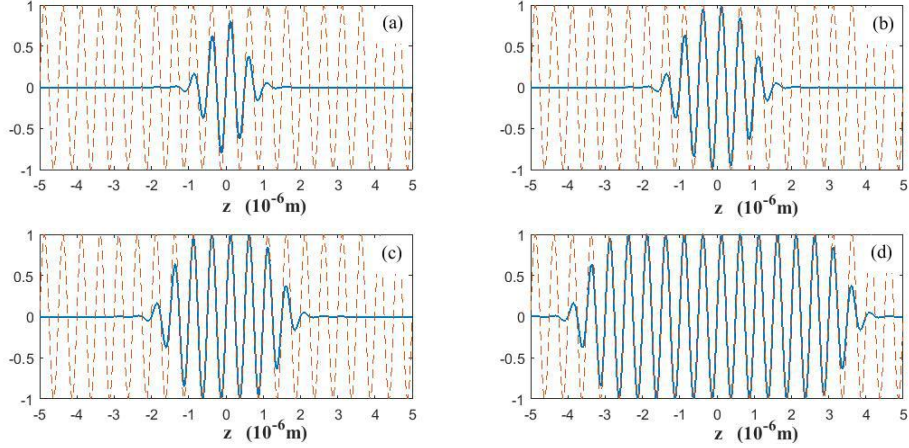

Fig. 1. Morlet wavelet reconstruction of classical plane wave with $E_{k0} = 1V/m$. The dotted line in the figure is the waveform of the plane light wave. Solid lines are the reconstructed light wave packets or light wave trains with discrete wavelet structure. where (a) -(d) are the cases for n = 0,1,2, and 6 with Eq. (10), respectively.

We have obtained the discrete wavelet structure of a plane light wave. Now we study the discrete energy of the light wave associated with its discrete wavelet structure.

For plane electromagnetic waves in free space, the energy density of them is $\varepsilon_0/2 \cdot |\vec{E}|^2 + 1/2\mu_0 \cdot |\vec{B}|^2 = \varepsilon_0 |\vec{E}|^2$, and $\vec{E} = \sum_k \vec{E}_k(\vec{r},t)$ is the total electric field of all possible plane electromagnetic waves. According to Eq.(3), In a finite volume $\tau = L_x \cdot L_y \cdot L_z$, the total energy of the plane electromagnetic waves is

$$\varepsilon_0 \int |\vec{E}|^2 d\tau = \varepsilon_0 \sum_{k,k'} \omega_k \omega_{k'} \int d\tau (q_k \vec{E}_{1k} + q_k^* \vec{E}_{1k}^*)(q_{k'} \vec{E}_{1k'} + q_{k'}^* \vec{E}_{1k'}^*) \quad , \quad (12)$$

where $\omega_k \vec{E}_{1k} = \vec{E}_k(\vec{r})$ (or $\omega_k \vec{E}_{1k0} = \vec{E}_{k0}(\vec{r})$). Besides, we know that the plane electromagnetic waves satisfy the following periodic conditions:

$$k_x = \frac{2\pi}{L_x}l, \; k_y = \frac{2\pi}{L_y}m, \; k_z = \frac{2\pi}{L_z}n, \; (l,n,m = 0,\pm 1,\pm 2,\pm 3,\ldots; \text{not all zero}) \quad , \quad (13)$$

where each set of $(l,m,n)$ corresponds to a wave vector, by which we can get the following expressions

$$\begin{aligned}\int d\tau (\vec{E}_{1k} \cdot \vec{E}_{1k'}^*) &= \int_0^{L_x}\int_0^{L_y}\int_0^{L_z} (\vec{E}_{1k0} \cdot \vec{E}_{1k'0}) e^{i[(k_x - k_x')x + (k_y - k_y')y + (k_z - k_z')z]} dxdydz \\ &= (\vec{E}_{1k0} \cdot \vec{E}_{1k'0}) L_x L_y L_z \delta_{k_x k_x'} \delta_{k_y k_y'} \delta_{k_z k_z'}\end{aligned} \quad , \quad (14)$$

$$\int d\tau(\vec{E}_{1k}^* \cdot \vec{E}_{1k'}) = \int d\tau(\vec{E}_{1k} \cdot \vec{E}_{1k'}^*) \quad . \tag{15}$$

And then
$$\varepsilon_0 \int |\vec{E}|^2 d\tau = \sum_\lambda \varepsilon_0 L_x L_y L_z [|\vec{E}_{1k0\parallel}|^2 + |\vec{E}_{1k0\perp}|^2] \omega_k^2 [q_k q_k^* + q_k^* q_k] , \tag{16}$$

where $|\vec{E}_{1k0\parallel}|^2$ and $|\vec{E}_{1k0\perp}|^2$ corresponds to two independent polarized components, respectively. By considering only one of components, and make the following transformation $Q_k = \varepsilon_0 \tau |\vec{E}_{1k0}|^2 (q_k + q_k^*)$, $\dot{Q}_k = \dot{q}_k + \dot{q}_k^* = -i\omega_k \varepsilon_0 \tau |\vec{E}_{1k0}|^2 (q_k - q_k^*)$, Eq.(16) can be expressed as

$$\varepsilon_0 \int |\vec{E}|^2 d\tau = \sum_k [\frac{1}{2}\dot{Q}_k^2 + \frac{1}{2}\omega_k^2 Q_k^2] = \sum_k H_k . \tag{17}$$

According to the classical canonical transformation, we can get that $H_k = \dot{Q}_k^2/2 + \omega_k^2 Q_k^2/2 = P_k \omega_k$ [22], where $P_k$ is a constant independent of time. For a given $\lambda$, we choose $L_x <\simeq \lambda$, $L_y <\simeq \lambda$ and $L_z = m\lambda, (m = 1, 2, 3, \ldots)$. According to periodic conditions above, we have $k_x = k_y = 0$, $k_z \neq 0$, which mean a plane wave propagating along the z-direction. By using the relation of $L_z = m\lambda$, the $P_k$ for this plane wave can be represented as $mP_k'$ since $\tau = L_x L_y L_z = m(L_x L_y \lambda)$. Then, by dropping the subscript of $\omega_k$, the energy for the plane wave in volume $\tau$ becomes

$$H_k = mP_k' \omega . \tag{18}$$

The result is true for any value of $m$. In other words, the energy of a segment of $m$ periods' electromagnetic wave in an infinitely long plane wave is proportional to the number of periods. This is consistent with the energy division of Planck radiation theory [23]. As is well known that, in a cavity of volume $V$, the number of electromagnetic vibration modes (oscillators) with frequency between $\nu$ and $\nu+d\nu$ is $N = 8\pi V \nu^2/c^3 \cdot d\nu$. And in his radiation theory, Planck supposed that the total energy belong to these $N$ oscillators can be divided into $P$ portions of $\varepsilon$ (i.e. the total energy is $P\varepsilon$), where $P$ is an integer. The energy $P\varepsilon$ is then assigned to $N$ oscillators statistically and the average energy per oscillator is $U = \varepsilon/(e^{\varepsilon/\kappa T} - 1) = h\nu/(e^{h\nu/\kappa T} - 1)$, where $h$, $\kappa$ and $T$ are the Planck's constant, Boltzmann constant and absolute temperature, respectively. So it is that $NU = 8\pi V/c^3 \cdot h\nu^3/(e^{h\nu/\kappa T} - 1) \cdot d\nu = Ph\nu$, i.e., $P = 8\pi V/c^3 \, hv^2/(e^{hv/\kappa T}-1) \cdot dv$ ; also $P/N = 1/(e^{hv/\kappa T}-1)$, from which we know that: ⅰ）When $hv/\kappa T \gg 1$, $P/N = e^{-hv/\kappa T} \ll 1$, which means that even if the energy is distributed evenly, each oscillator cannot get one portion of energy $hv$; ⅱ）$hv/\kappa T \ll 1$, $P/N = \kappa T/hv \gg 1$, which means that each oscillator can get far more than one of $hv$; and further $P = 8\pi V/c^3 \, \kappa T v/h \cdot dv$, i.e., $P \propto v \propto 1/\lambda$, meaning that the number of photons in given volume $V$ is inverse proportional to $\lambda$. This supports our result above, which is $m \propto 1/\lambda$ in given volume $\tau = m(L_x L_y \lambda)$.

A similar result for the wave train described by Eq. (11) can also be drawn by taking a segment of wave with complete periods. According to the above discussion on discrete wavelet structure of plane light wave, one can find that the wave train described by Eq.(11) is with $m = 2(n-1)$ complete periods, whose energy is then $2(n-1)P'_k \omega$ in volume $\tau$. And for the whole wave train in a volume $\tau'$ with cross-section $L_x \times L_y$ (where $L_x \lesssim \lambda$, $L_y \lesssim \lambda$), the numerical results show that it carries energy $0.59(2P'_k \omega)$ for $n=0$ or $(n+0.58)2P'_k \omega$ for $n \geq 1$. By denoting $2P'_k = p_k$, the energy in $\tau'$ carried by light wave of Eq. (11) can then be expressed as

$$H'_k = \begin{cases} 0.59 p_k \omega & (n=0) \\ (n+0.58) p_k \omega & (n=1,2,3\ldots < \infty) \end{cases} \quad . \tag{19}$$

A similar discussion can be made on a general cross-section for plane light wave packet or wave train. By letting $n \to \infty$, the result therefore becomes that for the infinite plane wave of Eq.(10).

Let's continue our discussion on the light wave described by Eq.(11). For a general value of $n$, if we regard it as a state of an admissive electromagnetic wave mode (degree of freedom) with idler frequency $\omega = kc$ rather than simply take it as a wave packet or wave train not indispensable, then the $E_{k0}$ in Eq.(11) could not be zero. Since, it is well known that electromagnetic vibration in free space does not depend on a medium. And the electromagnetic wave modes only depend on the electromagnetic vibrations. If we accept that each of admissive electromagnetic wave mode is a physical existence and always nonempty, then the electromagnetic vibration for its mode should always exists. If $E_{k0}=0$, the mode with it will disappear. So $E_{k0}$ would have a minimum value of nonzero，$E_{k0\min}$, which cannot be further divided. In other words, $p_k$ in Eq.(19) has a minimum value of nonzero, denoted by $p_{0k}$, being indivisible either, which suggests that a general $p_k$ is composed of some $p_{0k}$. For the case of $p_k = p_{0k}$, Eq.(19) reads

$$H'_{0k} = \begin{cases} 0.59 p_{0k}\omega & (n=0) \\ (n+0.58)p_{0k}\omega & (n=1,2,3...<\infty) \end{cases}. \qquad (20)$$

Therefore, the plane light wave packet or wave train Eq.(11) with $E_{k0\min}$ is the basic one. One sees that, for such kind of basic light wave packet with $n=1$, the minimum changeable energy is almost a portion of $p_{0k}\omega$ in an absorption process. the remaining part of $0.59 p_{0k}\omega$ could not be further absorbed for ensuring the existence of its electromagnetic wave vibration mode. Therefore, for a basic plane light wave packet or wave train described by Eq.(11), its changeable energy can be only with the form $H_{0k} = n p_{0k}\omega \ (n=1,2,3,...)$. Now there is a question: how to determine the minimum value of $p_{0k}$? The answer is experiment, for example that of photoelectric effect [24]. And the value of $p_{0k}$ is expected to be $h/2\pi$. One would also ask, whether the wave packet of $n=1$ can show the wavelength $\lambda (=2\pi/k_\lambda)$ of its electromagnetic wave mode in the experiment? We will study this by simulation of Mach-Zehnder interference of single photons [4].

We use interference field of random wave packets with $n=1$ and with minimum amplitude $E_{k0\min}$ to perform the simulation. The interference field of each pair of random wave packets can be expressed as

$$F(t,x,u) = \frac{E_{k0\min}}{2} e^{i\omega t} \sum_{j=-1}^{2} e^{-\frac{(t-jT)^2}{2s/c^2}} + \frac{E_{k0\min}}{2} e^{i\omega(t+\frac{x}{c})} \sum_{j=-1}^{2} e^{-\frac{(t+\frac{x}{c}+u-jT)^2}{2s/c^2}}, \qquad (21)$$

where $T = \lambda/c$ with $\lambda = 500 nm$; $x$ is the optical path difference between two arms of Mach-Zehnder interferometer; $u$ is a random time in the range $[-300T, 300T]$, which means that two wave packets leave the source with a random time difference $u$. For simplicity, we take $E_{k0\min}$ as two units. And the intensity distribution of the interference field after the superposition of $m$ pairs of wave packets is

$$I(x) = \sum_{n=1}^{m} \frac{1}{16T} \int_{-8T+u_{x,n}}^{8T+u_{x,n}} [\text{Re}(F(t,x,u_{x,n}))]^2 dt \qquad (22)$$

where $u_{x,n}$ is the random time corresponding to optical path difference $x$. Figure 2 shows the numerical results for $m = 3, 30, 300, 3000$, respectively. Figures.2 (a) and (b) are for the cases of $m = 3$ and $m = 30$, where none of interference fringe can be observed. When $m = 300$, see Fig.2(c), the interference fringes appear but not so clear. For a large enough number of $m$, for example, $m = 3000$, the interference fringes become very distinct as shown in Fig. 2(d). One can see that the space of the interference fringes (corresponding to the optical path difference) is exactly $500nm$. The result here is very like the experiment of Mach-Zehnder interference of single photons [4], and like the electron two-hole interference [25] as well. The light wave packets here behave like "photons" with "wave-particle duality". In the simulation, the wavelength of light can be revealed by numerous of wave packets through interference but cannot by a few of wave packets, which show none of wave information about light, however, behave as if they are "particles" like electrons [25].

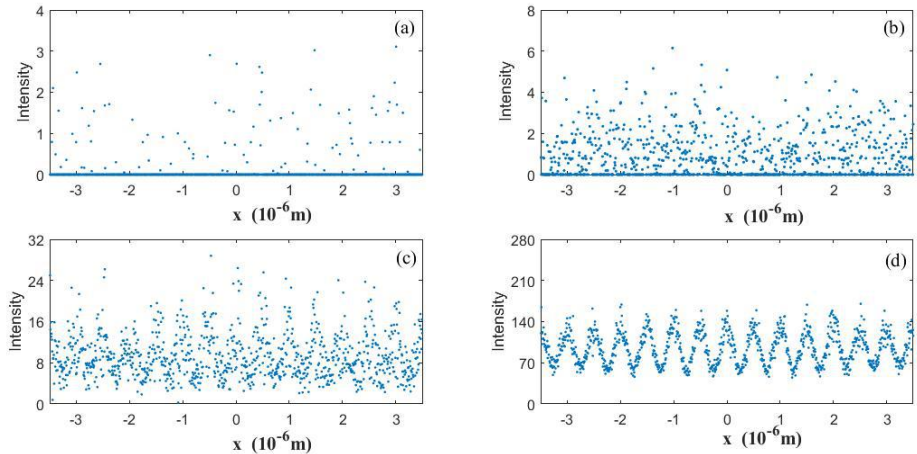

Fig. 2 Simulation on Mach-Zehnder interference of single photons by using the random light wave packets with discrete wavelet structure of $n=1$, where $m$ is the number of wave packet pairs for each optical path difference $x$; and (a) for $m = 3$, (b) for $m = 30$, (c) for $m = 300$, (d) for $m = 3000$, respectively.

In conclusion, we have used wavelet transform to decompose the classical linearly polarized plane light wave into a series of discrete Morlet wavelets or basic wavelets, and find out the relation between the energy and discrete wavelet structure of the light wave. The changeable energy of basic plane light wave packet or wave train is shown to be with the form of $H_{0k} = np_{0k}\omega \ (n = 1, 2, 3, ...)$. Finally, we make a simulation on the Mach-Zehnder interference of single photons by using the random light wave packets with discrete wavelet structure of $n=1$, showing the "wave-particle duality" of light.